# Doubly resonant photoacoustic spectroscopy: ultra-high sensitivity meets ultra-wide dynamic range

Zhen Wang[1,2], Qiang Wang[1,3,*], Hui Zhang[1,3], Simone Borri[4], Iacopo Galli[4], Angelo Sampaolo[5], Pietro Patimisco[5], Vincenzo Luigi Spagnolo[5], Paolo De Natale[4], and Wei Ren[2,*]

[1]State Key Laboratory of Applied Optics, Changchun Institute of Optics, Fine Mechanics and Physics, Chinese Academy of Sciences, Changchun 130033, China

[2]Department of Mechanical and Automation Engineering, The Chinese University of Hong Kong, New Territories, Hong Kong SAR, China

[3]University of Chinese Academy of Sciences, Beijing 100049, China

[4]CNR-INO – Istituto Nazionale di Ottica, and LENS – European Laboratory for Nonlinear Spectroscopy, 50019 Sesto Fiorentino, Italy

[5]PolySense Lab – Dipartimento Interateneo di Fisica, University and Politecnico of Bari, Via Amendola 173, Bari, Italy

Corresponding author names: Qiang Wang and Wei Ren

**Email:** wangqiang@ciomp.ac.cn; renwei@mae.cuhk.edu.hk

**Abstract**

Photoacoustic spectroscopy (PAS) based gas sensors with high sensitivity, wide dynamic range, low cost, and small footprint are desirable across a broad range of applications in energy, environment, safety, and public health. However, most works have focused on either acoustic resonator to enhance acoustic wave or optical resonator to enhance optical wave. Herein, we develop a gas sensor based on doubly resonant PAS in which the acoustic and optical waves are simultaneously enhanced using combined optical and acoustic resonators in a centimeter-long configuration. Not only the lower detection limit is enhanced by the double standing waves, but also the upper detection limit is expanded due to the short resonators. As an example, we developed a sensor by detecting acetylene ($C_2H_2$), achieving a noise equivalent absorption of $5.7\times10^{-13}$ cm$^{-1}$ and a dynamic range of eight orders. Compared to the state-of-the-art PAS gas sensors, the developed sensor increases the sensitivity by two orders of magnitude and extends the dynamic range by three orders of magnitude. Besides, a laser-cavity-molecule locking strategy is proposed to provide additional flexibility of fast gas detection.

**Introduction**

Laser-based optical gas sensors are increasingly required in many fields such as environmental monitoring (1,2), marine science (3,4), biological studies (5,6), and breath analysis (7,8). By measuring light attenuation through a gas sample, quantitative gas analysis can be performed. To enhance sensitivity, the absorption path length can be increased to hundreds of meters by a multipass cell or several kilometers by a high-finesse optical cavity. Among the most sensitive laser-based gas sensing techniques, noise-immune cavity-enhanced optical heterodyne molecular spectroscopy (NICE-OHMS) achieved $10^{-14}$ cm$^{-1}$ in noise equivalent absorption (NEA) (9). The minimum detectable concentration of a specific molecule has been achieved by saturated-absorption cavity ring-down (SCAR) with a limit of a few parts-per-quadrillion (ppq) (6,10). However, the relatively long cavity used in this class of gas sensors sets a limit on the dynamic range because the high finesse degrades even at low gas concentrations, which also leads to, in turn, the large overall footprint and weight. For example, with a cavity finesse of 30,000 and a cavity length of 37.8 cm used in NICE-OHMS, the dynamic range is limited to four orders of magnitude (11). Dynamic range is a very important parameter of gas sensors. For example, the $NH_3$ concentration changes by orders from tens of parts-per-million (ppm) to parts-per-billion (ppb) in applications including chemical hazards monitoring for occupational safety and health, contaminant control in hydrogen source for fuel cell, and contaminant requirement in lithography processing in semiconductor industry (12).

To get a combined high sensitivity and wide dynamic range, photoacoustic spectroscopy (PAS) is a promising candidate as the acoustic signal linearly increases with laser power, rather than a long absorption path. Numerous studies have been focused on applying various acoustic transducers and designing resonators to enhance the acoustic wave (13-16). An optical resonator can also be used to enhance the laser power by precisely matching the cavity mode and laser frequency using low-frequency dither locking (17,18), optical feedback (19,20), and Pound-Drever-Hall (PDH) locking method (21,22). However, compared to the well-assessed cavity ring-down spectroscopy (CRDS) and NICE-OHMS techniques ($10^{-14}$–$10^{-13}$ cm$^{-1}$ in NEA), the NEA of state-of-the-art PAS gas sensors (13,14,17,21–24) is limited to $10^{-11}$–$10^{-8}$ cm$^{-1}$, while the dynamic range is limited to five orders of magnitude. It is worth mentioning that researchers recently tried to combine an acoustic resonator with an optical resonator to advance the performance (25). However, the sensitivity and dynamic range cannot be simultaneously improved because of the limited optical and acoustic buildup factors, noise floor limited by the optical feedback strategy and the long optical cavity with a Brewster window (25). A high-efficiency doubly resonant strategy, which can enhance the intracavity PAS signal by orders linearly, is worth pursuing. Another limitation of the above techniques is the response time. To record the entire spectrum, the laser and optical cavity need to be relocked in a stepwise manner which is time-consuming (several minutes). The sensor will become much faster if the laser wavelength can be stabilized to the target absorption line while being locked with the optical cavity.

In this work, we report a PAS sensor with opto-acoustic resonance enhancement for gas detection with ultra-high sensitivity and ultra-wide dynamic range. The proposed photoacoustic design, leveraging on a double standing wave effect, achieves a combined acoustic amplification factor of 144 and laser power enhancement of almost three orders of magnitude. The laser-cavity-molecule locking strategy, which enables the simultaneous locking of laser frequency, cavity mode and absorption line, has proven a fast response. As a proof-of-principle, we show that our $C_2H_2$ sensor reaches a record sensitivity of $10^{-13}$ cm$^{-1}$ (NEA) and a record dynamic range of eight orders of magnitude.

**Results**

Figure 1A shows the schematic of opto-acoustic resonance for PAS. The core gas sensing element consists of an optical resonator, an acoustic resonator, and an acoustic transducer. Indeed, when the laser is in resonance with optical resonator via the laser-cavity locking, a standing optical wave is formed between the resonator mirrors. A high-finesse optical resonator can significantly build up the laser power by several orders of magnitude (26). PAS relies on the detection of acoustic waves generated by the periodic local heating and thermal expansion in the vibration-translation relaxation process of excited molecules after absorbing photons. To generate acoustic waves, the laser intensity is modulated at the resonant frequency of the acoustic resonator. A specifically designed one-dimensional longitudinal tube can be used to amplify the acoustic signal by forming a standing acoustic wave. In this work, we use both a quartz tuning fork (QTF) and an electret microphone to measure the PAS signal. The QTF locates nearby the antinode of the standing acoustic wave (27). The other configuration using an electret microphone is described in Supplementary Note 1. The photoacoustic signal ($S$) is given by (28):

$$S = b \times g \times K \times W_{in}(\lambda)\left(1 - e^{-\alpha_{eff}(\lambda)}\right)\varepsilon(f, \tau(P)) \tag{1}$$

where $b$ is the laser power buildup factor, $g$ is the acoustic wave enhancement factor, $K$ is the sensor constant, $W_{in}$ is the incident laser power, $\lambda$ is the laser wavelength, $\alpha_{eff}(\lambda)$ is the effective absorbance by the analyte, $\tau(P)$ is the relaxation time at the gas pressure $P$, and $\varepsilon$ is the radiation-to-sound conversion efficiency. Note that the factor $g$ is related to the geometry, to the material, and to the Q-factor of the acoustic resonator (29). The power buildup factor $b$ is determined by finesse of the optical resonator which needs to be selected properly (see Supplementary Note 2). Besides, a portion of the laser interrogates a low-pressure reference cell. The laser-molecule locking can be achieved by feedback to the cavity length. More details are provided in Supplementary Note 3.

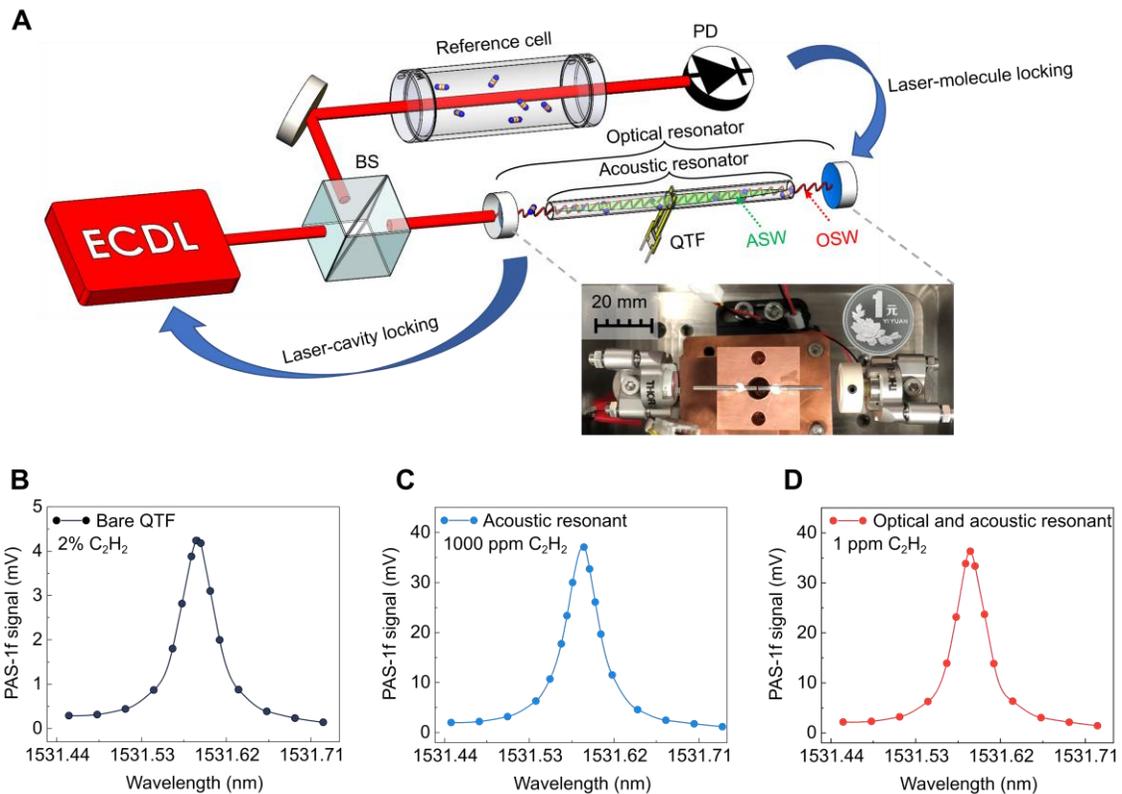

**Figure 1. (A)** Schematic representation of the experimental apparatus. ECDL: external cavity diode laser; BS: beam splitter; PD: photodetector; QTF: quartz tuning fork; ASW: acoustic standing wave; OSW: optical standing wave. The photo shows the real configuration by comparing with the 1 Chinese Yuan coin. PAS-1f signal measured by three different configurations: **(B)** bare QTF, **(C)** QTF with acoustic resonator, and **(D)** QTF with acoustic and optical resonators.

To evaluate the enhancement effects, the PAS-1f signal of the $C_2H_2$ line at 1531.59 nm is measured. Figure 1B, C and D compare the typical PAS-1f signals measured using a bare QTF (2% $C_2H_2$), a QTF with the mere acoustic resonator (0.1% $C_2H_2$), and a QTF with the complete opto-acoustic resonator (1 ppm $C_2H_2$). All the experiments are performed at the same incident laser power of 12 mW, lock-in detection bandwidth of 1 Hz, and gas pressure of 760 Torr. After normalization by the gas concentration, a comparison of Fig. 1B and Fig. 1C shows that the PAS-1f signal is enhanced by 175 times. However, with the high concentration $C_2H_2$ used in the bare QTF measurement, we must consider attenuation for the incident power before reaching the QTF. With a power attenuation of 17.3%, the enhancement factor of using the acoustic resonator is evaluated as 144. Besides, the optical resonator provides another enhancement factor of 980, as emerging from a comparison of Fig. 1C and Fig. 1D. Hence, the combined opto-acoustic amplification provides an overall enhancement of the PAS signal by a factor of $10^5$ via the double standing wave effect.

The measurement of PAS-1f signal is performed with 12-mW incident power at 760 Torr for different $C_2H_2$ concentrations (100 ppb and 10 ppb in nitrogen balance) and high-purity (99.999%) nitrogen when the laser wavelength is tuned from 1531.32 nm to 1531.75 nm, as shown from Fig. 2A to Fig. 2C. Note that the background signal is contributed by the thermoelastic effect due to unwanted absorption at the optical windows and resonator mirrors (30). A neighboring water line with quite low intensity of 2.896×10$^{−24}$ cm$^{−1}$/(molec·cm$^{−2}$) near 1531.37 nm is detected. This is probably due to residual water in the gas chamber. To reduce the measurement error and background, a multi-spectral fitting method based on HITRAN database (31), is implemented. Figure 2D compares the PAS-1f signal at 1 ppb $C_2H_2$ under two different incident powers. The PAS-1f amplitude is increased by a factor of about 22.5 when the incident power is increased from 12 mW to 300 mW (a factor of 25). The slight deviation of the enhancement factor between laser power and PAS signal is caused by the variation of the optical coupling efficiency. The linear relationship between signal-to-noise ratio and incident power is shown in Supplementary Note 4.

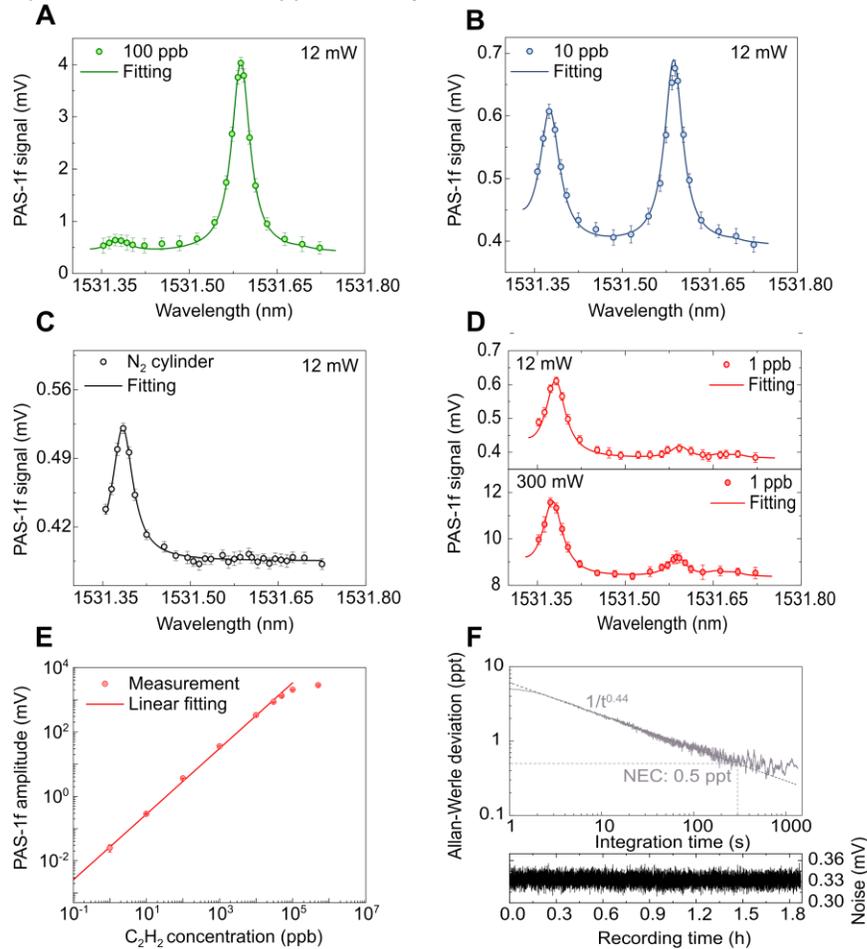

**Figure 2.** Representative PAS-1f signal of $C_2H_2/N_2$ mixtures and $N_2$ balance gas: **A** 100 ppb, **B** 10 ppb and **C** $N_2$ cylinder. **D** PAS-1f signal of 1 ppb $C_2H_2$ measured at two different incident optical power levels: 12 mW (top panel) and 300 mW (bottom panel). Due to the wide measurement span, error bars in **A, B, D**-top panel and **D**-bottom panel are magnified by 20, 2, 2 and 20 times, respectively, for the sake of clarity. **E** Background-subtracted PAS-1f amplitude as a function of gas concentration for an incident optical power of 12 mW. Error bars show the 1-σ standard deviation from 120 s measurements. **F** Allan–Werle deviation analysis. The dashed line in the top panel represents a $1/t^{0.44}$ slope. The bottom panel depicts the raw data of noise measured for nearly 2 hours.

Figure 2E shows the background-subtracted PAS-1f amplitude as a function of $C_2H_2$ concentration, from 1 ppb to 500 ppm. The sensor shows a very good linear response with a slope of 36.4 µV/ppb and an R-square value of 0.99 from 1 ppb to 50 ppm. Considering the 12-mW incident optical power, the normalized linear response coefficient of the sensor is 3 µV/(mW • ppb). Due to the verified linear relationship between incident optical power and PAS-1f amplitude, the response coefficient remains unchanged at 300 mW. However, the sensor deviates from the linear response at higher $C_2H_2$ concentrations due to the apparent degradation of finesse of the optical resonator. The Allan–Werle deviation analysis is conducted by tuning the laser wavelength to the peak of the $C_2H_2$ absorption line and measuring nitrogen, as shown in Fig. 2F. The noise equivalent concentration (NEC) is determined to be 5.1 ppt at an integration time of 1 s. At an incident optical power of 300 mW and a detection bandwidth of 1 Hz, we obtain an NNEA coefficient of $1.7 \times 10^{-12}$ W · cm$^{-1}$ · Hz$^{-1/2}$ (see Methods). The NEC can be improved to 0.5 ppt at a longer integration time of 300 s, leading to a NEA coefficient of $5.7 \times 10^{-13}$ cm$^{-1}$. As a result, the proposed gas sensor achieves a dynamic range of $1.0 \times 10^8$ (see Methods). Besides, the dynamic stability of the sensor is evaluated by operating the gas sensor when continuously filling $C_2H_2/N_2$ gas samples into the gas chamber (see Supplementary Note 5).

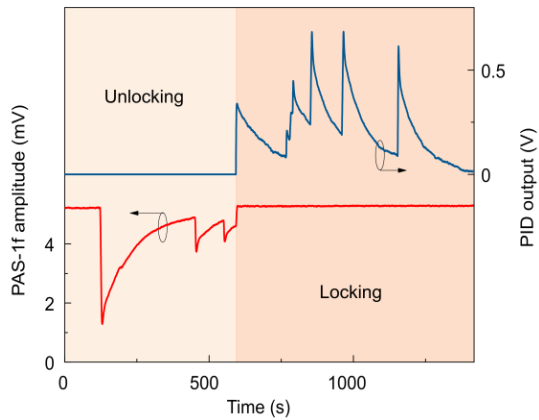

**Figure 3.** Continuous PAS measurement with and without locking the laser and molecular line. Unlocking state: laser and molecular line are not locked; locking state: laser and molecular line are locked. Note that the laser is kept being locked to the optical cavity during the entire test. The PID output refers to the laser-molecule locking loop.

To evaluate the performance of the laser-cavity-molecule locking strategy, we purposely apply external heating to the gas chamber to disturb the system. With 5 ppm $C_2H_2/N_2$ filled in the gas chamber, the laser wavelength is first tuned to the absorption line-center without implementing the molecular line-locking. Figure 3 shows that the PAS-1f amplitude varies sharply due to heat disturbance. This is because the laser wavelength drifts with the varied cavity length. Note that the gas temperature inside the chamber has little change during the heating disturbance (~ 3 s). Conversely, when the laser-cavity-molecule locking is activated, the PAS-1f amplitude remains unchanged under the condition of external heating disturbance. The PID output shows the compensation for the varied cavity length caused by the heating source. In this way, the laser, optical cavity and absorption line are tightly locked, leading to the immunity to laser wavelength drift and cavity length variation. The PAS-1f amplitude can thus be acquired continuously with no need for scanning the entire spectrum. Note that we use a relatively low gas pressure of 10 Torr in this test to avoid spectral interference.

**Discussion**
The key parameters of the state-of-the-art gas sensors are summarized in Table 1. Compared to PAS gas sensors, our work shows a dynamic range that is three orders of magnitude wider than the state-of-the-art ($10^5$) (23), and an NEA that is two orders of magnitude better than the state-of-the-art ($10^{-11}$ cm$^{-1}$) (22). Compared to other ultrasensitive spectroscopic methods such as NICE-OHMS, SCAR, CRDS and off-axis integrated cavity output spectroscopy (OA-ICOS) (6,9,32,33), NICE-OHMS and CRDS exhibit 57 times and 4 times better sensitivities, respectively. But our sensor shows evident merits of size (5–8 times shorter cavity) and dynamic range (2–3 orders wider). Besides, our sensor can operate at atmospheric pressure, which is more suitable for field applications.

**Table 1.** Key parameters of state-of-the-art gas sensors.

| Gas | Method | | λ(μm) | L(mm) | D | NEC(ppt) | NEA(cm⁻¹) | t(s) | P(Torr) |
|---|---|---|---|---|---|---|---|---|---|
| $NO_2$ | PAS (23) | P A S | 0.45 | NA | 1×10⁵ | 54 | 2.3×10⁻¹⁰ | 1 | 640 |
| $SF_6$ | QEPAS (24) | | 10.5 | NA | 3×10³ | 50 | 6.1×10⁻⁹ | 1 | 75 |
| $CO_2$ | I-QEPAS (17) | | 4.33 | 174 | 3×10³ | 300 | 1.4×10⁻⁸ | 20 | 38 |
| $C_2H_2$ | CECEPAS (22) | | 1.53 | 147 | 7×10³ | 24 | 2.4×10⁻¹¹ | 100 | 150 |
| $C_2H_2$ | CE-PAS (21) | | 1.53 | 130 | 1×10⁴ | 300 | 1.9×10⁻¹⁰ | 300 | 50 |
| HF | CEPAS (13) | | 2.47 | NA | 4×10⁴ | 0.65 | 1.3×10⁻¹⁰ | 1920 | 150 |
| $SF_6$ | α-$BiB_3O_6$-PAS (14) | | 10.6 | NA | 1×10⁵ | 0.75 | 8.1×10⁻¹¹ | 100 | 760 |
| $C_2H_2$ | **This work** | | 1.53 | 60 | 1×10⁸ | 0.5 | 5.7×10⁻¹³ | 300 | 760 |
| | | | | | 1×10⁷ | 5.1 | 5.8×10⁻¹² | 1 | |
| $^{14}CO_2$ | SCAR (6) | C E A S | 4.5 | 1000 | NS | 0.005 | 1.1×10⁻¹² | 7200 | 9.12 |
| $CO_2$ | CRDS (32) | | 1.6 | 330 | 7×10⁵ | 40000 | 1.4×10⁻¹³ | 4 | 0.075 |
| $C_2HD$ | NICE-OHMS (9) | | 1.06 | 469 | NS | NS | 1×10⁻¹⁴ | 1 | 0.0018 |
| CO | OA-ICOS (33) | | 1.57 | 1100 | NS | 1000 | 1.9×10⁻¹² | 1 | 197 |

λ: laser wavelength. L: cavity length. D: dynamic range. t: integration time. P: pressure. CEAS: cavity-enhanced absorption spectroscopy. NA: not applicable. NS: not stated.
Note that the dynamic range values are calculated using the same definition in the Methods. The two PAS sensors (13,14) that can achieve sub-ppt sensitivity are both based on mid-infrared lasers. The NEAs with different integration times are listed here because different sensors have different stability. A sensor may have a sensitivity improvement by averaging if it is stable enough with white noise dominant. The integration time is not normalized here as it cannot reflect the ultimate performance of sensors.

In summary, we have developed a photoacoustic gas sensor based on double standing wave enhancement. The photoacoustic signal has been enhanced by five orders of magnitude compared to a bare QTF, thus providing ultrasensitive gas detection. The optical resonator with a moderate finesse and a short length maintains the high optical coupling efficiency and enables a wide dynamic range. A strategy of laser-cavity-molecule locking is proposed to increase the sensor response for fast and continuous measurements. As a result, we have demonstrated an NEC of 0.5 ppt, an NEA of 5.7×10⁻¹³ cm⁻¹, and a dynamic range of eight orders of magnitude for $C_2H_2$ detection. Although the demonstration is performed in the near-infrared region, the concept can be extended to the mid-infrared range, where many molecules have much stronger fundamental absorption bands.

**Materials and Methods**

**NNEA calculation.** The NNEA coefficient can be calculated by the following equation:

$$NNEA = \frac{\alpha_{\min} \times W_{in}}{\sqrt{BW}}$$

where $\alpha_{min}$ is the NEA, $W_{in}$ is the incident power and $BW$ is the detection bandwidth. With the NEC of 5.1 ppt, NEA can be easily obtained based on the HITRAN database (31). Considering the incident power of 300 mW and the detection bandwidth of 1 Hz, the NNEA is 1.7×10⁻¹² W·cm⁻¹·Hz⁻¹/².

**Dynamic range definition.** The dynamic range is defined as the maximum concentration that is in the linear fitting range divided by the NEC (34–36). In this work, the maximum concentration in the linear fitting range is 50 ppm and the NEC is 0.5 ppt, leading to a dynamic range of eight orders of magnitude.

**References**

1. Kreuzer, L. B. & Patel, C. K. N. Nitric oxide air pollution: detection by optoacoustic spectroscopy. *Science* **173**, 45-47 (1971).
2. Rieker, G. B. et al. Frequency-comb-based remote sensing of greenhouse gases over kilometer air paths. *Optica* **1**, 290-298 (2014).
3. Kort, E. A. et al. Atmospheric observations of Arctic Ocean methane emissions up to 82º north. *Nat. Geosci.* **5**, 318-321 (2012).
4. Thornton, B. F. et al. Shipborne eddy covariance observations of methane fluxes constrain Arctic sea emissions. *Sci. Adv.* **6**, eaay7934 (2020).
5. Adato, R. & Altug, H. In-situ ultra-sensitive infrared absorption spectroscopy of biomolecule interactions in real time with plasmonic nanoantennas. *Nat. Commun.* **4**, 1-10 (2013).
6. Galli, I. et al. Spectroscopic detection of radiocarbon dioxide at parts-per-quadrillion sensitivity. *Optica* **3**, 385-388 (2016).
7. Wang, C. & Sahay, P. Breath analysis using laser spectroscopic techniques: breath biomarkers, spectral fingerprints, and detection limits. *Sensors* **9**, 8230-8262 (2009).
8. Liang, Q. et al. Ultrasensitive multispecies spectroscopic breath analysis for real-time health monitoring and diagnostics. *Proc. Natl. Acad. Sci.* USA **118**, e2105063118 (2021).
9. Ye, J. et al. J. L. Ultrasensitive detections in atomic and molecular physics: demonstration in molecular overtone spectroscopy. *J. Opt. Soc. Am. B* **15**, 6-15 (1998).
10. Delli Santi, M. G. et al. Biogenic Fraction Determination in Fuel Blends by Laser‐Based $^{14}CO_2$ Detection. *Adv. Photonics Res.* **2**, 2000069 (2021).
11. Foltynowicz, A. et al. Noise-immune cavity-enhanced optical heterodyne molecular spectroscopy: current status and future potential. *Appl. Phys. B* **92**, 313 (2008).
12. Manakasettharn, S. et al. Highly sensitive and exceptionally wide dynamic range detection of ammonia gas by indium hexacyanoferrate nanoparticles using FTIR spectroscopy. *Anal. Chem.* **90**, 4856-4862 (2018).
13. Tomberg, T. et al. Sub-parts-per-trillion level sensitivity in trace gas detection by cantilever-enhanced photo-acoustic spectroscopy. *Sci. Rep.* **8**, 1848 (2018)
14. Xiong, L. et al. Photoacoustic trace detection of gases at the parts-per-quadrillion level with a moving optical grating. *Proc. Natl Acad. Sci.* **114**, 7246-7249 (2017).
15. Wu, H. et al. Quartz enhanced photoacoustic $H_2S$ gas sensor based on a fiber-amplifier source and a custom tuning fork with large prong spacing. *Appl. Phys. Lett.* **107**, 111104 (2015).
16. Zheng, H. et al. Single-tube on-beam quartz-enhanced photoacoustic spectroscopy. *Opt. Lett.* **41**, 978-981 (2016).
17. Borri, S. et al. Intracavity quartz-enhanced photoacoustic sensor. *Appl. Phys. Lett.* **104**, 091114 (2014).
18. Wojtas, J. et al. Mid-infrared trace gas sensor technology based on intracavity quartz-enhanced photoacoustic spectroscopy. *Sensors,* **17**, 513 (2017).
19. Hippler, M. et al. Cavity-enhanced resonant photoacoustic spectroscopy with optical feedback cw diode lasers: A novel technique for ultratrace gas analysis and high-resolution spectroscopy. *J. Chem. Phys.* **133**, 044308 (2010).
20. Kachanov, A. et al. Cavity-enhanced optical feedback-assisted photo-acoustic spectroscopy with a 10.4 µm external cavity quantum cascade laser. *Appl. Phys. B* **110**, 47-56 (2013).
21. Wang, Z. et al. Ultrasensitive photoacoustic detection in a high-finesse cavity with Pound–Drever–Hall locking. *Opt. Lett.* **44**, 1924-1927 (2019).
22. Tomberg, T. et al. Cavity-enhanced cantilever-enhanced photo-acoustic spectroscopy. *Analyst* **144**, 2291-2296 (2019).
23. Yin, X. et al. Sub-ppb nitrogen dioxide detection with a large linear dynamic range by use of a differential photoacoustic cell and a 3.5 W blue multimode diode laser. *Sens. Actuat. B Chem.* **247**, 329-335 (2017).
24. Spagnolo, V. et al. Part-per-trillion level $SF_6$ detection using a quartz enhanced photoacoustic spectroscopy-based sensor with single-mode fiber-coupled quantum cascade laser excitation. *Opt. Lett.* **37**, 4461-4463 (2012).
25. Hayden, J. et al. Mid-Infrared intracavity quartz-enhanced photoacoustic spectroscopy with pptv–level sensitivity using a T-shaped custom tuning fork. *Photoacoustics*, **25**, 100330 (2022).

26. Friss, A. J. et al. Cavity-enhanced rotational Raman scattering in gases using a 20 mW near-infrared fiber laser. *Opt. Lett.* **41**, 3193-3196 (2016).
27. Dong, L. et al. QEPAS spectrophones: design, optimization, and performance. *Appl. Phys. B* **100**, 627-635 (2010).
28. Russo, S. D. et al. Quartz-enhanced photoacoustic spectroscopy exploiting low-frequency tuning forks as a tool to measure the vibrational relaxation rate in gas species. *Photoacoustics* **21**, 100227 (2021).
29. Miklós, A. et al. Application of acoustic resonators in photoacoustic trace gas analysis and metrology. *Rev. Sci. Instrum.* **72**, 1937-1955 (2001).
30. Liu, L. et al. Laser Induced Thermoelastic Contributions from Windows to Signal Background in a Photoacoustic Cell. *Photoacoustics* **22**, 100257 (2021).
31. Gordon, I. E. et al. The HITRAN2016 molecular spectroscopic database. *J. Quant. Spectrosc. Radiat. Transfer* **203**, 3-69 (2017).
32. Burkart, J. et al. Optical feedback frequency stabilized cavity ring-down spectroscopy. *Opt. Lett.* **39**, 4695-4698 (2014).
33. Engel, G. S. et al. Ultrasensitive near-infrared integrated cavity output spectroscopy technique for detection of CO at 1.57 μm: new sensitivity limits for absorption measurements in passive optical cavities. *Appl. Opt.* **45**, 9221-9229 (2006).
34. Jin, W. et al. Ultra-sensitive all-fibre photothermal spectroscopy with large dynamic range. *Nat. Commun.* **6**, 7767 (2015).
35. Zhao, P. et al. Mode-phase-difference photothermal spectroscopy for gas detection with an anti-resonant hollow-core optical fiber. *Nat. Commun.* **11**, 847 (2020).
36. Lou, X. et al. Ultra-wide-dynamic-range gas sensing by optical pathlength multiplexed absorption spectroscopy. *Photonics Res.* **9**, 193 (2021).

# Supplementary Information

## Supplementary Note 1: Sensor performance with a microphone

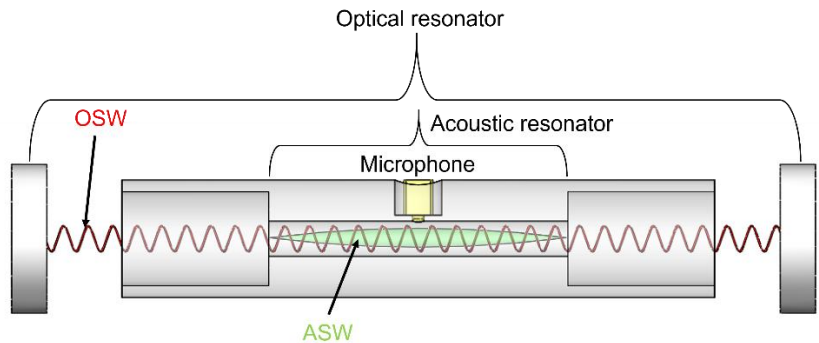

**Supplementary Figure 1.** Principle of the opto-acoustic resonance for PAS using a microphone. ASW: acoustic standing wave; OSW: optical standing wave. The amplitude of the OSW is modulated at the resonant frequency of the acoustic resonator. Two buffers are designed on both sides to form the acoustic standing wave with an antinode located at the center of the resonator.

The principle of the opto-acoustic resonance for PAS with a microphone as the acoustic transducer is illustrated in Supplementary Fig. 1. The cylindrical PAS cell is designed to have a total length of 70 mm, which is placed inside an optical resonator with a length of 80 mm. In particular, the acoustic resonator has a length of 35 mm and an inner diameter of 3 mm, and the two buffers on both sides of the acoustic resonator have a length of 17.5 mm and an inner diameter of 12 mm. The photoacoustic cell is characterized to have a resonant frequency of 3.5 kHz and a Q-factor of 25 at 760 Torr. An electret microphone is mounted at the center of the photoacoustic cell for acoustic detection, which corresponds to the antinode of the standing acoustic wave. The representative PAS-1f signals measured at different $C_2H_2$ concentrations (0.1–10 ppm) with an incident power of 12 mW are shown in Supplementary Fig. 2.

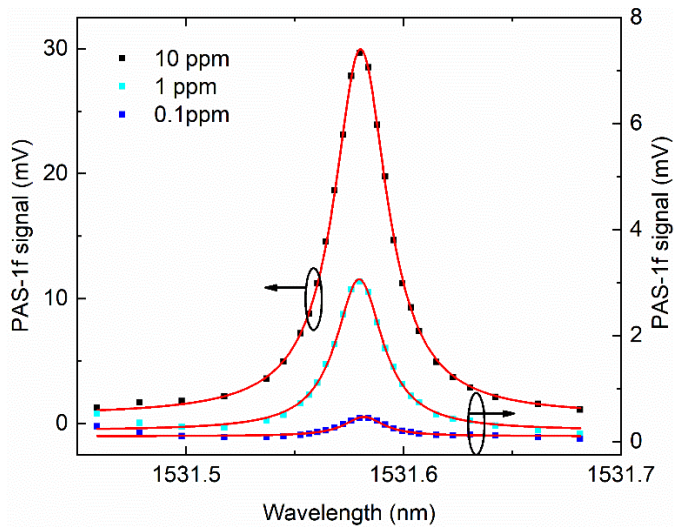

**Supplementary Figure 2.** Representative PAS-1f signals of $C_2H_2$ measured at different concentrations along with the spectral fitting curves at 760 Torr. The target $C_2H_2$ peak is observed at 1531.59 nm. The detection bandwidth of the lock-in amplifier is 1 Hz.

The linear response of the sensor is studied by measuring $C_2H_2$ of different concentrations shown in Supplementary Fig. 3. The sensor shows an excellent linear response from 1 ppb to 30 ppm with an R-squire value of 0.999. The Allan-Werle deviation analysis is plotted in Supplementary Fig. 4 by measuring nitrogen for about 1.5 hours at 760 Torr. The detection bandwidth of the lock-in amplifier is 1 Hz. As a result, the sensor can achieve a noise equivalent concentration (NEC) of 1.75 ppt at an integration time of 180 s, corresponding to a noise equivalent absorption (NEA) of $1.9 \times 10^{-12}$ cm$^{-1}$. In this case, the dynamic range of the sensor is about seven orders of magnitude.

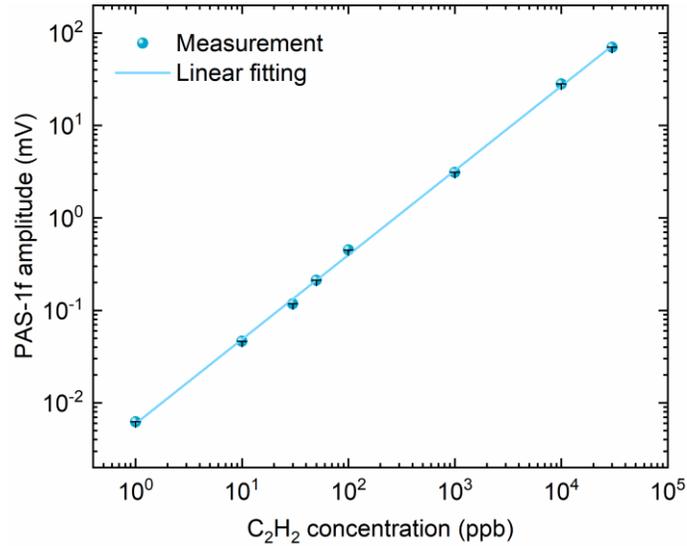

**Supplementary Figure 3.** Background-subtracted PAS-1f amplitude as a function of $C_2H_2$ concentration. The linear fitting yields an R-square value of 0.999 from 1 ppb to 30 ppm. The incident laser power is 12 mW and the detection bandwidth of the lock-in amplifier is 1 Hz. Error bars show the 1-σ standard deviation for 120 s measurements.

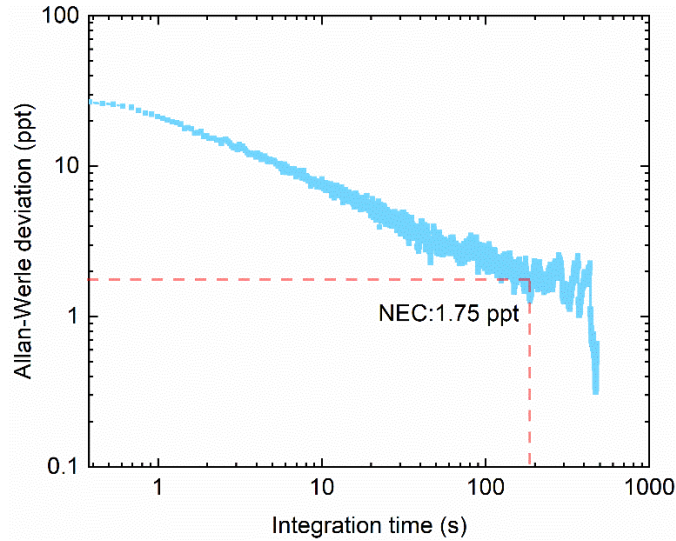

**Supplementary Figure 4.** Allan-Werle deviation analysis for the microphone-based sensor. The laser wavelength is tuned to the peak of the $C_2H_2$ absorption line with the chamber filled with nitrogen. As a function of integration time, the analysis shows a NEC of 1.75 ppt at 180 s. The detection bandwidth is 1 Hz, the same as signal measurement.

## Supplementary Note 2: Selection of finesse for the optical resonator

First, the selection of a higher finesse theoretically provides a larger power enhancement factor. However, the linewidth of the cavity mode decreases with the increased finesse (1), making it more challenging to achieve a high coupling efficiency of the laser into the optical resonator. Second, the dynamic range of the sensor is affected by the finesse. The finesse is determined by the round-trip loss inside the optical resonator which includes the mirror transmission and absorption, gas absorption, and loss induced by the acoustic module (1). The finesse increases with the larger reflectivity of cavity mirrors, but it decreases apparently when the loss induced by gas absorption becomes comparable (1–3). Hence, the physical length of the optical resonator should be selected as short as possible to minimize the gas-absorption loss inside the cavity.

# Supplementary Note 3: Experimental setup in detail

## 3.1 Photoacoustic sensor apparatus

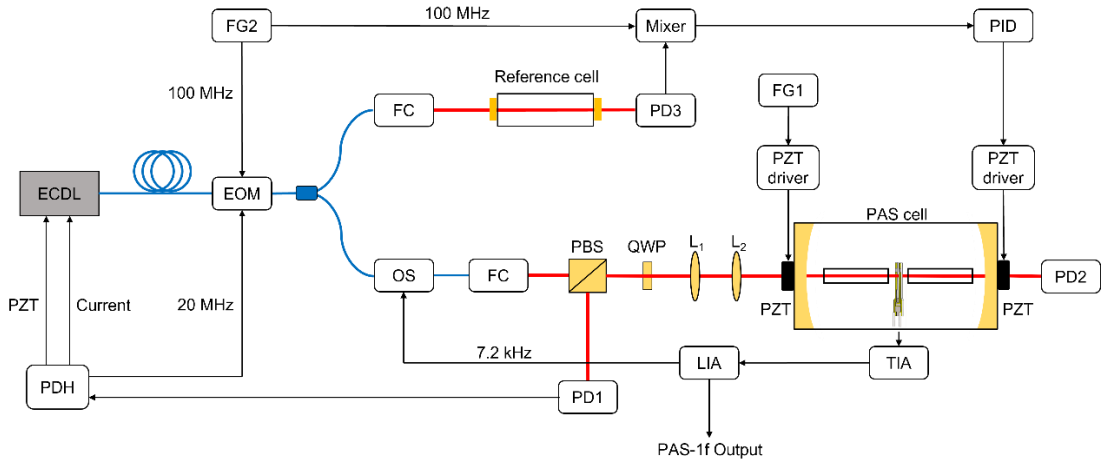

**Supplementary Figure 5.** Schematic of the photoacoustic sensor with opto-acoustic resonance. ECDL, external cavity diode laser; EOM, electro-optic modulator; OS, optical switch; FC, fiber collimator; PBS, polarization beam splitter; QWP, quarter-wave plate; PD, photodetector; PZT, piezo transducer; FG, function generator; PDH, phase demodulator and laser servo (PDD110/mFALC110, TOPTICA Photonics); PID, proportion integration differentiation controller (LB1005, Newport); TIA, trans-impedance amplifier; LIA, lock-in amplifier.

The schematic of the doubly resonant PAS sensor is shown in Supplementary Fig. 5. An external cavity diode laser (ECDL, TOPTICA Photonics) is used to detect the P(11) transition of $C_2H_2$ at 1531.59 nm. The ECDL is phase modulated by an electro-optic modulator (EOM, iXblue Photonics) at 20 MHz and locked to the optical resonator using the PDH method (4). The polarization beam splitter is used with the quarter-wave plate to pick up the reflected beam from the optical resonator, which is then detected by a photodetector (PD1). Details of the application of the PDH method to PAS can be found in our previous work (3). In this work, the current and piezo transducer (PZT) feedback loops of the ECDL are both used to improve the locking performance. Each mirror (Layertec Inc.) of the optical resonator has a radius of curvature of 150 mm and reflectivity of 99.923% (finesse 4078) at the laser wavelength, as measured by cavity ring-down (see Supplementary Note 3.2). Compared with other cavity-enhanced absorption spectroscopic methods (5,6), the optical resonator used here for the opto-acoustic resonance features a much shorter length (60 mm in this work). Two mode matching lenses (focus length: $L_1$=30 mm and $L_2$=50 mm) and a photodetector (PD2) are used to maximize the coupling efficiency (84%) between the laser and the optical resonator. With a maximum incident laser power of 300 mW, the intracavity optical power is boosted to 264 W in this work (see Supplementary Note 3.3).

The intracavity laser beam passes through the acoustic resonator, consisting of two stainless-steel tubes (inner diameter 1.3 mm, length 23 mm), and does not touch any surface. The central axis of the acoustic resonator is about 1.2 mm below the top of the QTF prongs, thus optimizing the piezoelectrical conversion efficiency (7). The two tubes are placed at a distance of ~60 μm from the QTF, so that it lies near the antinode of the acoustic wave and leaves its Q-factor unaffected. The beam waist (340 μm in diameter) is located between the two prongs of the QTF, which has a gap of 800 μm, a resonant frequency of 7.2 kHz and a Q-factor of ~8000 (gas pressure 760 Torr) (7). The optical resonator, the acoustic resonator and the QTF are all enclosed inside a chamber (PAS cell in Supplementary Fig. 5). A high-speed lithium niobate optical switch (NanoSpeed, Agiltron) is used to chop the laser beam at the same frequency as the resonant frequency of the QTF. The piezoelectric current from the QTF is collected by a trans-impedance amplifier and then amplified by a low-noise voltage preamplifier (SR560, Stanford Research Systems). Finally, a lock-in amplifier (MFLI 5 MHz, Zurich Instruments) with a detection bandwidth of 1 Hz is used to demodulate the first harmonic signal (1f) at the sensor output. To retrieve a complete spectrum, the optical cavity length is tuned by a PZT attached to one cavity mirror via the function generation FG1.

Besides the laser-cavity locking using the PDH method, another pair of sidebands are generated by applying a 100 MHz modulation (FG2) to the same EOM, if locking the laser to the absorption line is also required. A small portion of the incident laser passes through a 3-m reference gas cell filled by 2% $C_2H_2$ (pressure 7.6 Torr) and impinges on a photodetector PD3. The error signal is retrieved by mixing the PD3 signal with a 100 MHz reference signal, which is used to PID-control the other cavity mirror that is attached to a PZT. As a result, these two feedback loops enable the simultaneous locking of laser frequency, cavity mode and absorption line. More details of this laser-cavity-molecule locking strategy will be described in Supplementary Note 3.4.

### 3.2 Characterization of mirror reflectivity

To characterize the mirror reflectivity using ring-down measurement, a longer optical resonator of 600 mm is designed with two high-reflectivity mirrors which have a radius of curvature (ROC) of 1500 mm. Note that all the mirrors used in this study are fabricated in the same coating process and have the same reflectivity. An optical switch with 85-ns fall time is used to swiftly generate the ring-down event, which is monitored by a high-bandwidth (150 MHz) photodetector. The ring-down signal, illustrated in Supplementary Fig. 6, is acquired by a high-speed data acquisition card with a 100-MHz sampling rate and a 16-bit analog-to-digital converter.

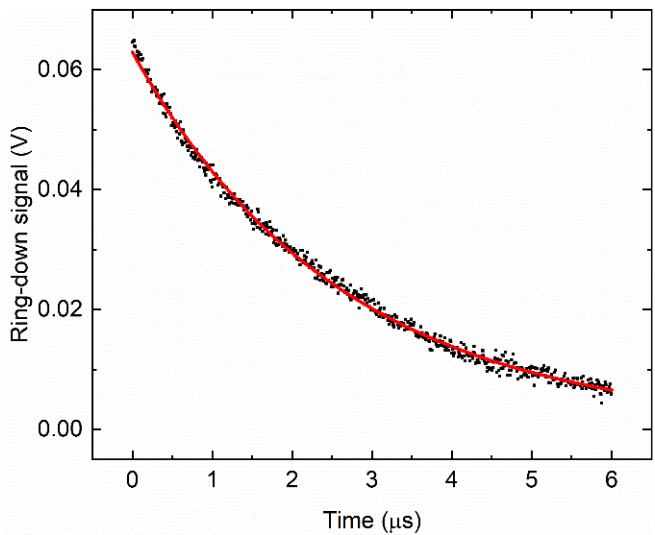

**Supplementary Figure 6.** Characterization of mirror reflectivity using cavity ring-down measurement. The averaged ring-down time (average of 84 ring-down events) is determined to be 2.596 µs, corresponding to a reflectivity of 99.923%.

### 3.3 Optical coupling efficiency and intracavity power

It is important to couple as much light as possible into the optical resonator. To evaluate the coupling efficiency, the reflected light from the optical resonator is detected by a photodetector. By scanning the resonator length using a piezo transducer (PZT), the fundamental resonator mode shown in Supplementary Fig. 7 is used to investigate the coupling efficiency. Based on the ground voltage (4.451 V) and valley voltage (0.719 V), a coupling efficiency of about 84% is determined in this work.

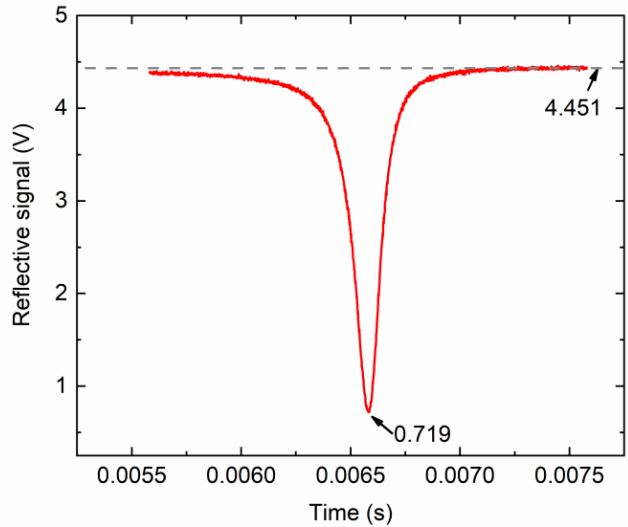

**Supplementary Figure 7.** Measured intensity of the reflected light from the optical resonator. A coupling efficiency of about 84% is determined from the ground voltage (4.451 V) and valley voltage (0.719 V).

According to the main text, the optical resonator enhances the photoacoustic signal by a factor of 980. Considering the 84% coupling efficiency, the intracavity power enhancement factor is 1166, which is slightly lower than the theoretical buildup factor of 1300. This is probably due to the extra cavity loss caused by the QTF. As the empty cavity has a finesse of 4078, an additional loss of 0.017% can deteriorate the buildup factor from 1300 to 1166. When the incident power is adjusted from 12 mW to 300 mW, there is an extra 10% degradation of coupling efficiency. Hence, the maximum of the intracavity power is determined to be 264 W, by the product of the incident laser power (300 mW), the power enhancement factor of the empty cavity (1166) and the coupling efficiency (84%×(1−10%)=75.6%).

### 3.4 Laser-cavity-molecule locking

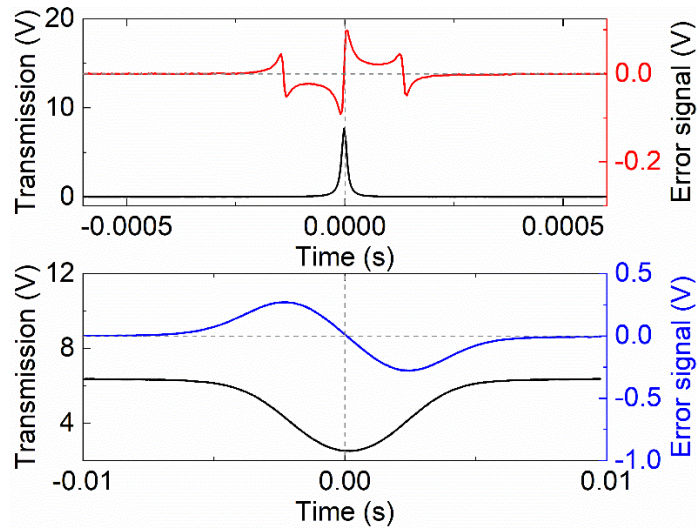

**Supplementary Figure 8.** Molecular line locking measurement. Typical transmission (black curve) and error signal (red and blue curves) measured during the laser-cavity locking (upper panel) and laser-molecule locking (bottom panel), respectively.

Rather than scanning the entire spectrum, it is of interest to explore the feasibility of the continuous photoacoustic measurement at the absorption line center. By scanning the laser wavelength, Supplementary Fig. 8 shows the representative transmission and the error signals for the laser-cavity locking and laser-molecule locking, respectively. In the upper panel of Supplementary Fig. 8, the center zero-point of the PDH error signal corresponds to the peak of cavity transmission. In the bottom panel of Fig. 8, the center zero-point corresponds to the absorption peak of the gas (2% $C_2H_2$, 7.6 Torr) in the reference gas cell.

## Supplementary Note 4: Signal-to-noise ratio with incident power

By repeating the measurement at 100 ppb $C_2H_2$, supplementary Fig. 9 shows the variation of the PAS-1f amplitude with the incident optical power. The sensor signal increases almost linearly (0.25 mV/mW) with the incident power. In contrast, the noise level remains almost unchanged over the entire power range, as shown in Supplementary Fig. 9 (1-$\sigma$ standard deviation of $N_2$ over 120 s). This makes our set-up very promising to further enhance detection sensitivity by simply increasing the incident laser power.

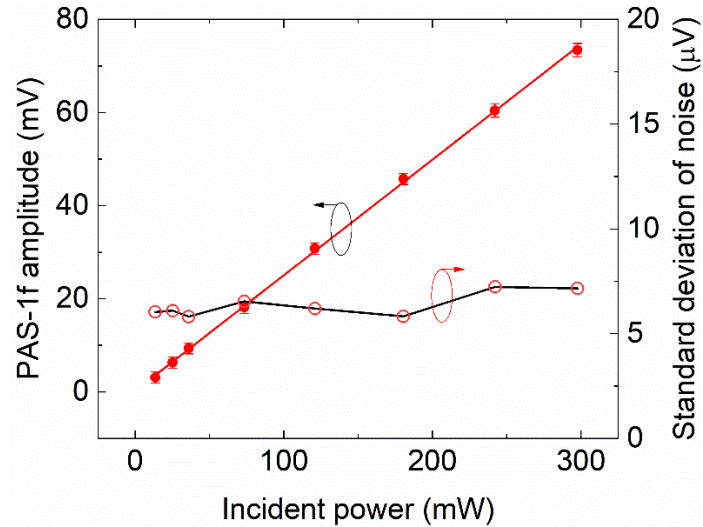

**Supplementary Figure 9.** PAS-1f amplitude for 100 ppb $C_2H_2$ and 1-$\sigma$ standard deviation (STD) of the noise (a measurement of nitrogen over 120 s) versus incident power. The PAS-1f amplitudes of 1 ppb $C_2H_2$ at different incident power levels need to be inferred from the fitting results. All these measurements are performed at 760 Torr and a detection bandwidth of 1 Hz. The error bars (1-$\sigma$ STD) are calculated from the raw data taken in a time interval of 120 s. Error bars are magnified by 200 times, for the sake of clarity.

## Supplementary Note 5: Continuous measurement

With the laser wavelength tuned at the peak of the absorption line, the sensor is tested by continuously filling 100 ppb $C_2H_2$ into the gas chamber. The flow rate is varied (200, 300, and 400 mL/min) at 60 s and 120 s, respectively, during the continuous measurement. As shown in Supplementary Fig. 10, the sensor can work continuously without losing locking due to the robust locking between laser and optical resonator.

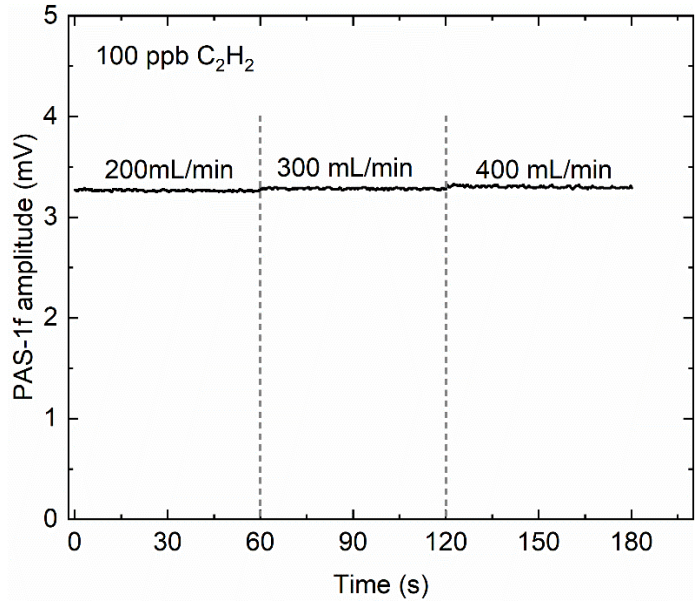

**Supplementary Figure 10.** Continuous monitoring of 100 ppb $C_2H_2$ at varied flow rates (200, 300, and 400 mL/min). The flow rate is changed at 60 s and 120 s, respectively, during the continuous measurement.

## Supplementary References

1. Romanini, D., Ventrillard, I., Méjean, G., Morville, J., & Kerstel, E. Introduction to Cavity Enhanced Absorption Spectroscopy. (Springer, 2014), pp.1-60.

2. Borri, S. et al. Intracavity quartz-enhanced photoacoustic sensor. Appl. Phys. Lett. 104, 091114 (2014).

3. Wang, Z. et al. Ultrasensitive photoacoustic detection in a high-finesse cavity with Pound–Drever–Hall locking. Opt. Lett. 44, 1924-1927 (2019).

4. Drever, R. W. P. et al. Laser phase and frequency stabilization using an optical resonator. Appl. Phys. B 31, 97-105 (1983).

5. Galli, I. et al. Spectroscopic detection of radiocarbon dioxide at parts-per-quadrillion sensitivity. Optica 3, 385-388 (2016).

6. Ye, J., Ma, L. S. & Hall, J. L. Ultrasensitive detections in atomic and molecular physics: demonstration in molecular overtone spectroscopy. J. Opt. Soc. Am. B 15, 6-15 (1998).

7. Wu, H. et al. Quartz enhanced photoacoustic $H_2S$ gas sensor based on a fiber-amplifier source and a custom tuning fork with large prong spacing. Appl. Phys. Lett. 107, 111104 (2015).